\documentclass{article}
\usepackage{makeidx}
\usepackage{graphicx}
\usepackage[binary-units=true]{siunitx}
\usepackage[numbers]{natbib}
\usepackage{listings}

\usepackage{amsmath}
\usepackage{amsthm}
\usepackage{todonotes}
\usepackage{amsopn}
\usepackage{algorithm}
\usepackage[noend]{algorithmic}
\usepackage{tikz}
\usepackage{tabularx}
\usepackage{subfig}
\usepackage[bookmarks=false]{hyperref}
\usetikzlibrary{positioning,shapes.geometric,calc,automata,arrows}
\usepackage{multirow}
\newcolumntype{Y}{>{\centering\arraybackslash}X}

\begin{document}

\title{Complementing Model Learning with Mutation-Based Fuzzing}

\author{Rick Smetsers$^1$ \and Joshua Moerman$^1$ \and Mark Janssen$^2$ \and Sicco Verwer$^2$}

\date{%
    $^1$ Radboud University Nijmegen\\ 
    \texttt{\{rick, moerman\}@cs.ru.nl}\\[2ex]
    $^2$ Delft University of Technology\\
    \texttt{mark@ch.tudelft.nl, s.e.verwer@tudelft.nl}
    }

\maketitle

\begin{abstract}
An ongoing challenge for learning algorithms formulated in the Minimally Adequate Teacher framework is to efficiently obtain counterexamples.
In this paper we compare and combine conformance testing and mutation-based fuzzing methods for obtaining counterexamples when learning finite state machine models for the reactive software systems of the Rigorous Exampination of Reactive Systems (RERS) challenge.
We have found that for the LTL problems of the challenge the fuzzer provided an independent confirmation that the learning process had been successful, since no  additional counterexamples were found.
For the reachability problems of the challenge, however, the fuzzer discovered more reachable error states than the learner and tester, albeit in some cases the learner and tester found some that were not discovered by the fuzzer.
This leads us to believe that these orthogonal approaches are complementary in the context of model learning.
\end{abstract}

\section{Introduction}

Software systems are becoming increasingly complex.
\emph{Model learning} is quickly becoming a popular technique for reverse engineering such systems.
Instead of viewing a system via its internal structure, model learning algorithms construct a formal model from observations of a system's behaviour.

One prominent approach for model learning is described in a seminal paper by \citet{angluin1987learning}.
In this work, she proved that one can effectively learn a model that describes the behaviour of a system if a so-called \emph{Minimally Adequate Teacher} is available.
This teacher is assumed to answer two types of questions about the (to the learner unknown) \emph{target}:
\begin{itemize}
\item In a \emph{membership query} (MQ) the learner asks for the system's output in response to a sequence of inputs.
The learner uses the outputs for a set of such queries to construct a \emph{hypothesis}.
\item In an \emph{equivalence query} (EQ) the learner asks if its hypothesis is equivalent to the target. 
If this is not the case, the teacher provides a \emph{counterexample}, which is an input sequence that distinguishes the hypothesis and the target.
The learner then uses this counterexample to refine its hypothesis through a series of membership queries.
\end{itemize}
This process iterates until the learner's hypothesis is equivalent to the target.

\citet{Peled1999} have recognized the avail of Angluin's work for learning models of real-world, reactive systems that can be modeled by a finite state machine (FSM).
Membership queries, on the one hand, are implemented simply by interacting with the system. 
Equivalence queries, on the other hand, require a more elaborate approach, as there is no trivial way of implementing them.
Therefore, an ongoing challenge, and the topic of this paper, is to efficiently obtain counterexamples.

\medskip

\noindent Several techniques for obtaining counterexamples have been proposed.
The most widely studied approach for this purpose is \emph{conformance testing} \citep{Dorofeeva2010}.
In the context of learning, the goal of conformance testing is to establish an equivalence relation between the current hypothesis and the target.  
This is done by posing a set of so-called \emph{test queries} to the system.
In a test query, similarly to a membership query, the learner asks for the system's response to a sequence of inputs. 
If the system's response is the same as the predicted response (by the hypothesis) for all test queries, then the hypothesis is assumed to be equivalent to the target.
Otherwise, if there is a test for which the target and the hypothesis produce different outputs, then this input sequence can be used as a counterexample.

One of the main advantages of using conformance testing is that it can distinguish the hypothesis from all other finite state machines of size at most $m$, where $m$ is a user-selected bound on the number of states.
This means that if we know a bound $m$ for the size of the system we learn, we are guaranteed to find a counterexample if there exists one.
Unfortunately, conformance testing has some notable drawbacks.
First, it is hard (or even impossible) in practice to determine an upper-bound on the number of states of the system's target FSM.
Second, it is known that testing becomes exponentially more expensive for higher values of $m$ \citep{Vasilevskii1973}. 
Therefore, the learner might incorrectly assume that its hypothesis is correct. 
This motivates the search for alternative techniques for implementing equivalence queries.

\medskip

\noindent The field of \emph{mutation-based fuzzing} provides opportunities here.
In essence, \emph{fuzzers} are programs that apply a test (i.e.\ input sequence) to a target program, and then iteratively modify this sequence to monitor whether or not something interesting happens (e.g.\ crash, different output, increased code coverage \ldots).
Fuzzers are mostly used for security purposes, as a crash could uncover an exploitable buffer overflow, for example. 
\emph{Mutation-based} fuzzers randomly replace or append  some inputs to the test query.

Recently, good results have been achieved by combining mutation-based fuzzing with a genetic (evolutionary) algorithm. 
This requires a fitness function to evaluate the performance of newly generated test query, i.e.\ a measurement of `how interesting' it is. 
In our case, this fitness function is based on what code is executed for a certain test query.
The fittest test cases can then be used as a source for mutation-based fuzzing.
Hence, tests are mutated to see if the coverage of the program is increased.
Iterating this process creates an evolutionary approach which proves to be very effective for various applications \citep{afl-website}.

\subsubsection*{RERS Challenge 2016}

\noindent In this report we describe our experiments in which we apply the aforementioned techniques to the \emph{Rigorous Examination of Reactive Systems} (RERS) challenge 2016.
The RERS challenge consists of two parts: 
\begin{enumerate}
\item \emph{problems} (i.e. reactive software) for which one has to prove or disprove certain logical properties, and 
\item problems for which one has to find the reachable error states.
\end{enumerate}

\noindent In our approach, we have used a state-of-the-art learning algorithm (\emph{learner}) in combination with a conformance testing algorithm (\emph{tester}) to learn models for the RERS 2016 problems.
In addition, we have used a mutation-based fuzzing tool (\emph{fuzzer}) to generate potentially interesting traces independently of the learner and the tester. 
We have used these traces as a verification for the learned models and found that
\begin{itemize}
\item For part (1) of the challenge the fuzzer did not find any additional counterexamples for the learner, compared to those found by the tester. Therefore the fuzzer provided an independent confirmation that the learning process had been successful.
\item For part (2) of the challenge the fuzzer discovered more reachable error states than the learner and tester, albeit in some cases the learner and tester found some that were not discovered by the fuzzer.
\end{itemize}

\noindent Our experiments lead us to believe that in some applications, fuzzing is a viable technique for finding \emph{additional} counterexamples for a learning setup.
In this report, in addition to describing our experimental setup for RERS in detail, we therefore describe possible ways of combining learning and fuzzing.

\section{Preliminaries}
In this section, we describe preliminaries on finite state machines, model learning, conformance testing, and fuzzing.

\subsection{Finite state machines}
A \emph{finite state machine} (FSM) is a model of computation that can be used to design computer programs.
At any time, a FSM is in one of its (finite number of) states, called the \emph{current state}.
Generally, the current state of a computer program is determined by the contents of the memory locations (i.e.\ variables) that it currently has access to, and the values of its registers, in particular the program counter.
Changes in state are triggered by an event or condition, and are called \emph{transitions}.
We assume that transitions are triggered based on events, or \emph{inputs}, that can be observed. 

\medskip

\noindent Formally, we define a FSM as a Mealy machine $M = (I, O, Q, q_M, \delta, \lambda)$, where $I, O$ and $Q$ are finite sets of \emph{inputs}, \emph{outputs} and \emph{states} respectively, $q_M \in Q$ is the \emph{start state}, $\delta: Q \times I \rightarrow Q$ is a \emph{transition function}, and $\lambda: Q \times I \rightarrow O$ is an \emph{output function}.
The functions $\delta$ and $\lambda$ are naturally extended to $\delta: Q \times I^{\ast} \to Q$ and $\lambda: Q \times I^{\ast} \to O^{\ast}$.
Observe that a FSM is deterministic and input-enabled (i.e.\ complete) by definition.

For $q \in Q$, we use $\lfloor q \rfloor_M$ to denote a \emph{representative} access sequence of $q$, i.e.\ $\delta(q_M, \lfloor q \rfloor_M) = q$.
We extend this notation to arbitrary sequences, allowing to transform them into representative access sequences: for $x \in I^{\ast}$, we define $\lfloor x \rfloor_M = \lfloor \delta(q_M, x)\rfloor_M$.

A \emph{discriminator} for a pair of states $q, q'$ is an input sequence $x \in I^{\ast}$ such that $\lambda(q, x) \neq \lambda(q', x)$ \cite{Smetsers2016}.

The behaviour of a FSM $M$ is defined by a \emph{characterization function} $A_M : I^{\ast} \to O^{\ast}$ with $A_M(x) = \lambda(q_M, x)$ for $x \in I^{\ast}$. 
FSMs $M$ and $M'$ are \emph{equivalent} if $A_M(x) = A_{M'}(x)$ for $x \in I^{\ast}$.

\subsection{Model learning} \label{sec:learning}
The goal of so-called active model learning algorithms is to learn a FSM $H = (I, O, Q_H, q_H, \delta_H, \lambda_H)$ for a system whose behaviour can be characterized by a (unknown) FSM $M = (I, O, Q_M, q_M, \delta_M, \lambda_M)$, given the set of inputs $I$ and access to the characterization function $A_M$ of $M$.

\medskip 

The \emph{TTT algorithm} is a novel model learning algorithm formulated in the MAT framework \cite{Isberner2014a}. 
The distinguishing characteristic of TTT is its redundancy-free handling of counterexamples.
The TTT algorithm maintains a prefix-closed set $S$ of access sequences to states.
These states correspond to leaves of a \emph{discrimination tree} $T$, in which the inner nodes are labeled with elements from a suffix-closed set of discriminators $E$, and its transitions are labeled with an output.

A hypothesis is constructed by \emph{sifting} the sequences in $S \cdot I$ through the discrimination tree: 
Given a prefix $ua$, with $u \in S$ and $a \in I$, starting at the root of $T$, at each inner node labelled with a discriminator $v \in E$ a membership query $A_M(uav)$ is posed. 
Depending on the last output of this query, we move on to the respective child of the inner node. 
This process is repeated until a leaf is reached.
The state in the label of the leaf becomes the target for transition $\delta(\delta(q_H, u),a)$.

\medskip

\noindent The way that the TTT algorithm handles counterexamples is based on the observation by \citet{Rivest1994} that a counterexample $x \in I^{\ast}$ can be decomposed in a prefix $u \in I^{\ast}$, input $a \in I$, and suffix $v \in I^{\ast}$ such that $x = uav$ and $A_M(\lfloor u \rfloor_H av) \neq A_M(\lfloor ua \rfloor_H v)$.
Such a decomposition shows that the state $q = \delta_H(\delta_H(q_H, u), a)$ is incorrect, and that this transition should instead point to a new state $q'$ with access sequence $\lfloor u \rfloor_H a$.
Therefore, this sequence is added to $S$.
Observe that this does not affect the prefix-closedness of $S$.
In the discrimination tree $T$, the leaf corresponding to $q$ is replaced with an inner node labelled by the \emph{temporary} discriminator $v$.
A technique known as \emph{discriminator finalization} is applied to construct the subtree of this newly created inner node, and obtain a minimal discriminator for $q$ and $q'$.
For a description of discriminator finalization, we refer to \cite{Isberner2014a}.


\subsection{Conformance testing} \label{sec:testing}
Conformance testing for FSMs is an efficient way of finding counterexamples.
Let $H = (I, O, Q_H, q_H, \delta_H, \lambda_H)$ be a hypothesis with $n$ states.
We call a conformance testing method $m$-complete if it can identify the hypothesis in the set of all FSMs with at most $m$ states.
Such $m$-complete methods are generally polynomially in the size of the hypothesis and exponential in $m - n$, which are far more efficient than an exhaustive search.
For an overview of some $m$-complete methods, we refer to \citet{Dorofeeva2010}.
All of these methods require the following information:
\begin{itemize}
\item A set of \emph{access sequences} $S = \{\lfloor q \rfloor_H | q \in Q_H\}$, possibly extended to a \emph{transition cover} set $S \cdot I$.
\item A \emph{traversal set} $I^{l}$ that contains all input sequences of length $l = m - n + 1$, where $m = |Q_M|$ and $n = |Q_H|
$.
\item A means of pairwise distinguishing all states of $H$, such as set of \emph{discriminators} $E$ for all pairs of states in $H$.
\end{itemize}
A test suite is then constructed by combining these sets, or subsets of these sets, e.g.\ $S \cdot I^{l} \cdot E$.
The difference between different testing methods is how states are distinguished (i.e.\ the last part).

In the so-called \emph{partial W-method}, or \emph{Wp-method}, \citep{Fujiwara1991} states are distinguished pairwise:
For each state $q \in Q_H$ a set $E_{q} \subset E$ of discriminators is constructed, such that for each state $q' \in Q \setminus \{q\}$ there is a sequence $w \in E_{q}$ that distinguishes $q$ and $q'$, i.e.\ $\lambda_H(q, w) \neq \lambda_H(q', w)$.
Then, each trace $uv, u \in S \cdot I, v \in I^{l}$ is extended with the set $E_q$ where $q = \delta_H(q_H, uv)$.



\medskip

\noindent Conformance testing is typically expensive due to the exponential size of the traversal set. 
Given a hypothesis $H$ with $n$ states and $k$ inputs, the worst-case length of a test suite (i.e. the sum of the length of all sequences) is of order $\mathcal{O}(k^{l}n^3)$ (recall that $l = m - n + 1$, where $m$ is the upper bound on the number of states of $M$).
Moreover, it is hard to estimate an upper bound for $M$ in practice. 


%

\subsection{Fuzzing} \label{sec:fuzzing}
A \emph{mutation-based fuzzer} is a program that applies a set of tests (i.e.\ input sequences) to a target program, and then iteratively mutates these tests to monitor if `something interesting' happens.
This could be a crash of the target program, a change in its output, or it finds that more code is covered (via instrumentation).
The \emph{American Fuzzy Lop} (AFL) fuzzer \citep{afl-website} is interesting for its approach in combining mutation-based test case generation with \emph{code coverage} monitoring.

AFL supports programs written in C, C++, or Objective C and there are variants that allow to fuzz programs written in Python, Go, Rust or OCaml.
AFL works on instrumented binaries of these programs, and supports compile-time or runtime instrumentation.
The tool is bundled with a modified version of gcc (afl-gcc) that can add instrumentation at compile time.
The compile-time instrumentation has the best performance, but requires the source code of the target program to be available. 
When the source code is not available, AFL applies runtime instrumentation, which uses emulation (QEMU or Intel Pin) to achieve the result. 
This, however, is 2-5{$\times$} slower than compile-time instrumentation \citep{afl-website}.

%

From a high-level, simplified perspective, AFL works by taking a program and a queue of tests, and iteratively mutating these tests to see if the \emph{coverage} of the program is increased; new tests that increase coverage are added to the queue. 
In the next paragraphs, we will describe in more detail how coverage is measured by AFL, which mutation strategies are applied, and how execution time is minimized.


\subsubsection*{Measuring coverage}
If a mutated test case results in a higher coverage of the target program, the test case is seen as valuable. 

In order to measure this coverage, AFL uses instrumentation of the control flow of the program (branches, jumps, etc.), to identify which parts of the target program are used in a given test. 
Using this knowledge, AFL can decide which test cases cover behaviour not previously seen in other test cases, simply by comparing the result of the instrumentation. 

Internally, coverage is measured by using a so-called \emph{trace bitmap}, which is a \SI{64}{\kilo\byte} array of memory shared between the fuzzer and the instrumented target. 
This array is updated by the following code every time an edge in the control flow is taken.
\begin{lstlisting}[language=C]
cur_location = <COMPILE_TIME_RANDOM>;
shared_mem[cur_location ^ prev_location]++;
prev_location = cur_location >> 1;
\end{lstlisting}
Every location in the array is represented by a compile-time random value. 
When an edge in the control flow is taken, the bitmap is updated at the position of the current location and an xor of the previous location value. 
The intention is that every edge in the control flow is mapped to a different byte in the bitmap. 

Note that because the size of the bitmap is finite and the values that represent locations in the code are random, the bitmap is probabilistic: there is a chance that collisions will occur. 
This is especially the case when the bitmap fills up, which can happen when fuzzing large programs with many edges in their control flow. 
AFL can detect and resolve this situation by applying instrumentation on fewer edges in the target or by increasing the size of the bitmap.


\subsubsection*{Mutation strategies}
At the core of AFL is its `engine' to generate new test cases. 
As mentioned earlier, AFL uses a collection of techniques to mutate existing test cases into new ones, starting with basic deterministic techniques and progressing onto more complex ones. 
The author of AFL has described the following strategies \citep{afl-blog-mutation}:
\begin{itemize}
	\item Performing sequential, ordered \emph{bit flips} to a sequence of one, two, or four bits of the input.
    \item An extension of bit flips to (a sequence of one, two or four) \emph{bytes}.
    \item Applying \emph{simple arithmetic} (incrementing and decrementing) to integers in the input.
    \item \emph{Overwriting} integers in the input by values from set of pre-set integers (such as -1, 1024 and \lstinline!MAX_INT!), that are known to trigger edge conditions in many programs.
    \item When the deterministic strategies (above) are exhausted, randomised \emph{stacked operations} can be applied, i.e.\ a sequence of single-bit flips, setting discovered byte values, addition and subtraction, inserting new random single-byte sets, deletion of blocks, duplication of blocks through overwrite or insertion, and zeroing blocks.
    \item The last-resort strategy involves taking two known inputs from the queue that cover different code paths and \emph{splicing} them in a random location. 
\end{itemize}

\subsubsection*{Fork server}
In general, fuzzers generate a lot of tests.
Therefore, many invocations of the target process are required.
Instead of starting a new process for every test, AFL uses a \emph{fork server} to speed up fuzzing.
The fork server initialises the target process only once, and then forks (clones) it to create a new instance for each test case.

On modern operating systems, a process fork is done in a copy-on-write fashion, which means that any memory allocated by the process is only copied when it is modified by the new instance. 
This eliminates most (slow) memory operations compared to a regular process start \citep{afl-blog-forkserver}, and allows for an execution of approximately \SI{10000} tests per second on a single core of our machine.

\section{Experimental Setup}

\noindent In this section we describe the experiments in which we apply the aforementioned techniques to the \emph{Rigorous Examination of Reactive Systems} (RERS) challenge 2016.
The RERS challenge consists of two parts: 
\begin{enumerate}
\item A set of nine \emph{problems} (i.e. reactive software), numbered 1 through 9, for which one has to prove or disprove a set of given \emph{linear temproal logic} (LTL) formulae, and 
\item a set of nine problems, numbered 10 through 18, for which one has to determine whether or not a set of error statements present in the source code are \emph{reachable}, and provide a sequence of inputs such that the error statement is executed.
\end{enumerate}

\noindent In our approach, we have used a state-of-the-art learner in combination with a tester to learn FSMs for the RERS 2016 problems.
In addition, we have used a fuzzer to generate potentially interesting traces independently of the learner and the tester. 
As we have executed the learner/tester and the fuzzer indepenedently of one another, we describe their experimental setup and result in turn.
The code for our experiments is available at \url{https://gitlab.science.ru.nl/moerman/rers-2016/}.

\subsection{Learning and Testing with LearnLib} \label{sec:exp_learning}
For our learning and testing experiments, we have used LearnLib, an open-source Java library for active model learning \citep{merten2011learnlib}.
As a learner in LearnLib consideres its system under learning as a black-box, we have interfaced LearnLib with a compiled binary of each of the 18 problems.
Below, we list and explain the choices we have made regarding our LearnLib setup.
\begin{description}
\item[Learning algorithm] For our learning algorithm, we have chosen the TTT algoritm as implemented in LearnLib, because previous experiments have shown that it scales up to larger systems under learning; both in the amount of membership queries asked and in the amount of memory used in the process. 
\item[Testing algorithm] For our testing algorithm, we have used our own implementation of the Wp method. 
Recall that the Wp method in principle generates a test suite whose size is polynomial in the size of the hypothesis and exponential in the upper bound of states in the system, minus the size of the hypothesis. 
Instead of exhausting this test suite, our implementation of the algorithm randomly samples test sequences until it finds a counterexample:
First, it samples a prefix uniformly from the state-cover set of the current hypothesis.
Then, it randomly generates an infix over all inputs according to a geometric distribution.
Finally, we sample a suffix uniformly from the set of state-specific discriminators.
By using a geometric distribution for the infix, we are not bounding the length of the test sequence.
In our tool, the minimal and expected length of the infix can be set by parameters.
In our experiments, its minimal length was three, and its expected length was eleven.
\item[Counterexample handling] Counterexamples were processed using the~\textsc{LinearForward} handler in LearnLib.
\item[Cache] We have used the cache that is implemented in LearnLib to avoid sending duplicate queries to the system under learning.
\end{description}

\noindent The final hypothesis for each of the problems was stored as a DOT file.
In order to solve the LTL formulae for part (1) of the challenge, these DOT files were translated to NuSMV.
For part (2), it sufficed to \texttt{grep} the DOT files for the unique outputs that were generated in an error state.

\subsection{Fuzzing with AFL} \label{sec:exp_fuzzing}
Independently of learning and testing the challenge's problems, we have used AFL to fuzz them.
Below, we give an overview of some of the details of our experimental setup.
\begin{description}
\item[Instrumentation] We have used the \texttt{afl-gcc} compiler that comes bundled with AFL to compile the C source code for each of the problems.
This compiler instruments the control flow of the program, and implements the fork server.
\item[Input alphabet] AFL requires an input alphabet as a source for its mutation strategies. We have used the valid inputs that were defined in the source code for each problem as an input alphabet.
\item[Error handling] In order to compile the reachability problems (10 - 18) an external error handling function had to be provided. 
This function is called with a unique identifier whenever an error state is reached.
Our implementation of the error function prints the unique error identifier, and then aborts the program.
This way, each trace whose execution leads to an error state is registered by AFL as a crash.
As these traces are stored in a separate results folder by AFL, we could easily separate them from traces that did not lead to an error state.
\item[Post-processing] As the input bytes that AFL considers are not limited to the valid inputs for the challenge problems, we filtered out the bytes that were not accepted.
\end{description}

\noindent The traces that were found by AFL were simulated on the final hypothesis of the learner to see if its output differed from that of the program binary.

\section{Results}

The results for the learning/testing setup described in \autoref{sec:exp_learning} are shown in \autoref{tab:learn_ltl} and \autoref{tab:learn_reach}.

We are confident that the learned models for the LTL problems (1--9) are complete, as the last hypothesis was learnt within 1 day and no further counterexamples were found in the following week. 
The same holds for the first of the reachability problems (10).
Beware, however, that we can never guarantee completeness with black-box techniques.

For problems 11--18 we know that we do \emph{not} have complete models, as the learner was still finding new states every 10 minutes when the server rebooted for maintenance.
The learner ran for a bit more than 7 days and saved all hypotheses.
As a result of this reboot, we do not have statistics on the number of queries.

\begin{table}
    \centering
	\caption{Learning and testing results for the LTL problems of RERS 2016 on an Intel(R) Xeon(R) CPU E7-4870 v2 @ 2.30GHz (server), with Oracle Java 8 JVM configured with a 40GB heap.}
	\label{tab:learn_ltl}
    {\renewcommand{\arraystretch}{1.5}
	\begin{tabular}{ l | l  l  l }
    size & plain & arithmetic & data structures \\
    \hline
    small & \parbox[t]{0.25\textwidth}{\textbf{Problem 1}\\time: 50s\\states: 13\\hypotheses: 6} & \parbox[t]{0.25\textwidth}{\textbf{Problem 2}\\time: 1m22s\\states: 22\\hypotheses: 10} & \parbox[t]{0.25\textwidth}{\textbf{Problem 3}\\time: 7m05s\\states: 26\\hypotheses: 13} \\[37pt]
    medium & \parbox[t]{0.25\textwidth}{\textbf{Problem 4}\\time: 34m\\states: 157\\hypotheses: 77} & \parbox[t]{0.25\textwidth}{\textbf{Problem 5}\\time: 2h43m\\states: 121\\hypotheses: 50} & \parbox[t]{0.25\textwidth}{\textbf{Problem 6}\\time: 4h51m\\states: 238\\hypotheses: 156} \\[37pt]
    large & \parbox[t]{0.25\textwidth}{\textbf{Problem 7}\\time: 11h45m\\states: 610\\hypotheses: 407} & \parbox[t]{0.25\textwidth}{\textbf{Problem 8}\\time: 24h22m\\states: 646\\hypotheses: 432} & \parbox[t]{0.25\textwidth}{\textbf{Problem 9}\\time: 18h31m\\states: 854\\hypotheses: 550} \\
    \end{tabular}
    }
\end{table}

\begin{table}
    \centering
	\caption{Learning and testing results for the reachability problems of RERS 2016 on an Intel(R) Xeon(R) CPU E7-4870 v2 @ 2.30GHz (server), with Oracle Java 8 JVM configured with a 40GB heap.}
	\label{tab:learn_reach}
    {\renewcommand{\arraystretch}{1.5}
	\begin{tabular}{ l | l  l  l }
    size & plain & arithmetic & data structures \\
    \hline
    small & \parbox[t]{0.25\textwidth}{\textbf{Problem 10}\\time: 2m39s\\states: 59\\hypotheses: 3} & \parbox[t]{0.25\textwidth}{\textbf{Problem 11}\\time: 1w+\\states: 22 589\\hypotheses: 8 314} & \parbox[t]{0.25\textwidth}{\textbf{Problem 12}\\time: 1w+\\states: 12 771\\hypotheses: 4 325} \\[37pt]
    medium & \parbox[t]{0.25\textwidth}{\textbf{Problem 13}\\time: 1w+\\states: 12 848\\hypotheses: 5 564} & \parbox[t]{0.25\textwidth}{\textbf{Problem 14}\\time: 1w+\\states: 11 632\\hypotheses: 4 513} & \parbox[t]{0.25\textwidth}{\textbf{Problem 15}\\time: 1w+\\states: 7 821\\hypotheses: 3 792} \\[37pt]
    large & \parbox[t]{0.25\textwidth}{\textbf{Problem 16}\\time: 1w+\\states: 8 425\\hypotheses: 3 865} & \parbox[t]{0.25\textwidth}{\textbf{Problem 17}\\time: 1w+\\states: 11 758\\hypotheses: 5 584} & \parbox[t]{0.25\textwidth}{\textbf{Problem 18}\\time: 1w+\\states: 8 863\\hypotheses: 4 246} \\
    \end{tabular}
    }
\end{table}

The results for the fuzzing setup described in \autoref{sec:exp_fuzzing} are shown in \autoref{tab:fuzz_ltl} and \autoref{tab:fuzz_reach}.
These results should be interpreted as follows:
\begin{description}
\item[cycles] The number of times the fuzzer went over all the interesting test
traces discovered, fuzzed them, and looped back to the very beginning.
\item[execs] The total number of test traces executed.
\item[paths] The total number of test traces found that have a unique execution path.
\end{description}

\begin{table}
    \centering
	\caption{Fuzzing results for the LTL problems of RERS 2016 on a Intel(R) Xeon(R) CPU E7-4870 v2 @ 2.30GHz (server). The fuzzer was terminated after approximately 10 days.}
	\label{tab:fuzz_ltl}
    {\renewcommand{\arraystretch}{1.5}
	\begin{tabular}{ l | l  l  l }
    size & plain & arithmetic & data structures \\
    \hline
    small & \parbox[t]{0.25\textwidth}{\textbf{Problem 1}\\cycles: 46 521\\execs: $2.64 \times 10^9$\\paths: 253} & \parbox[t]{0.25\textwidth}{\textbf{Problem 2}\\cycles: 30 088\\execs: $2.68 \times 10^9$\\paths: 480} & \parbox[t]{0.25\textwidth}{\textbf{Problem 3}\\cycles: 19 551\\execs: $2.52 \times 10^9$\\paths: 453} \\[37pt]
    medium & \parbox[t]{0.25\textwidth}{\textbf{Problem 4}\\cycles: 460\\execs: $8.68 \times 10^8$\\paths: 3 453} & \parbox[t]{0.25\textwidth}{\textbf{Problem 5}\\cycles: 4 191\\execs: $1.63 \times 10^9$\\paths: 1 115} & \parbox[t]{0.25\textwidth}{\textbf{Problem 6}\\cycles: 109\\execs: $7.32 \times 10^8$\\paths: 4 494} \\[37pt]
    large & \parbox[t]{0.25\textwidth}{\textbf{Problem 7}\\cycles: 35\\execs: $6.86 \times 10^8$\\paths: 9 556} & \parbox[t]{0.25\textwidth}{\textbf{Problem 8}\\cycles: 16\\execs: $6.75 \times 10^8$\\paths: 10 906} & \parbox[t]{0.25\textwidth}{\textbf{Problem 9}\\cycles: 71\\execs: $7.63 \times 10^8$\\paths: 11 305} \\
    \end{tabular}
    }
\end{table}

\begin{table}
    \centering
	\caption{Fuzzing results for the reachability problems of RERS 2016 on a Intel(R) Xeon(R) CPU E7-4870 v2 @ 2.30GHz (server). The fuzzer was terminated after approximately 10 days.}
	\label{tab:fuzz_reach}
    {\renewcommand{\arraystretch}{1.5}
	\begin{tabular}{ l | l  l  l }
    size & plain & arithmetic & data structures \\
    \hline
    small & \parbox[t]{0.25\textwidth}{\textbf{Problem 10}\\cycles: 70 336\\execs: $2.58 \times 10^9$\\paths: 139} & \parbox[t]{0.25\textwidth}{\textbf{Problem 11}\\cycles: 10 365\\execs: $2.34 \times 10^9$\\paths: 801} & \parbox[t]{0.25\textwidth}{\textbf{Problem 12}\\cycles: 5 971\\execs: $2.14 \times 10^9$\\paths: 1 032} \\[37pt]
    medium & \parbox[t]{0.25\textwidth}{\textbf{Problem 13}\\cycles: 779\\execs: $1.35 \times 10^9$\\paths: 4 235} & \parbox[t]{0.25\textwidth}{\textbf{Problem 14}\\cycles: 621\\execs: $1.02 \times 10^9$\\paths: 3 838} & \parbox[t]{0.25\textwidth}{\textbf{Problem 15}\\cycles: 1 040\\execs: $1.77 \times 10^9$\\paths: 3 685} \\[37pt]
    large & \parbox[t]{0.25\textwidth}{\textbf{Problem 16}\\cycles: 50\\execs: $7.22 \times 10^8$\\paths: 11 908} & \parbox[t]{0.25\textwidth}{\textbf{Problem 17}\\cycles: 19\\execs: $4.58 \times 10^8$\\paths: 10 283} & \parbox[t]{0.25\textwidth}{\textbf{Problem 18}\\cycles: 21\\execs: $4.58 \times 10^8$\\paths: 10 237} \\
    \end{tabular}
    }
\end{table}

For the LTL problems of the challenge, none of the test traces that have a unique execution path were counterexamples for the last hypothesis of the learner.
This, in combination with the large number of cycles completed by the fuzzer, strengthens our belief that the learned models for these problems (1 - 9) are complete.

The number of reachable error states found by the learner and the fuzzer are shown in \autoref{tab:errors}.
The first entry in each cell is the number of unique error states that were found, and the second entry is the number of error states that were found by the given technique, but were not found by the other technique (e.g. ``fuzzing: 28 (2)'' means that the fuzzer has found 28 error states, and 2 of those were not found by the learner).  

\begin{table}
    \centering
	\caption{Number of error states found.}
	\label{tab:errors}
    {\renewcommand{\arraystretch}{1.5}
	\begin{tabular}{ l | l  l  l }
    size & plain & arithmetic & data structures \\
    \hline
    small & \parbox[t]{0.25\textwidth}{\textbf{Problem 10}\\learner: 45 (0)\\fuzzer: 45 (0)\\total: 45} & \parbox[t]{0.25\textwidth}{\textbf{Problem 11}\\learner: 20 (0)\\fuzzer: 22 (2)\\total: 22} & \parbox[t]{0.25\textwidth}{\textbf{Problem 12}\\learner: 21 (0)\\fuzzer: 21 (0)\\total: 21} \\
    medium & \parbox[t]{0.25\textwidth}{\textbf{Problem 13}\\learner: 28 (0)\\fuzzer: 30 (2)\\total: 30} & \parbox[t]{0.25\textwidth}{\textbf{Problem 14}\\learner: 27 (0)\\fuzzer: 30 (3)\\total: 30} & \parbox[t]{0.25\textwidth}{\textbf{Problem 15}\\learner: 27 (0)\\fuzzer: 32 (5)\\total: 32} \\
    large & \parbox[t]{0.25\textwidth}{\textbf{Problem 16}\\learner: 29 (1)\\fuzzer: 31 (3)\\total: 32} & \parbox[t]{0.25\textwidth}{\textbf{Problem 17}\\learner: 27 (1)\\fuzzer: 28 (2)\\total: 29} & \parbox[t]{0.25\textwidth}{\textbf{Problem 18}\\learner: 28 (0)\\fuzzer: 32 (4)\\total: 32} \\
    \end{tabular}
    }
\end{table}

From these results we conclude that the fuzzer discovered more reachable error states than the learner/tester, albeit in some cases the learner/tester found some that were not discovered by the fuzzer.

\section{Present and Future Work}
The goal of our present and future research in this area is to combine model learning and mutation-based fuzzing in the following ways.
\begin{enumerate}
\item use fuzzing as a source of counterexamples \emph{during} learning, and
\item use (intermediate) learning results to guide mutation-based fuzzing.
\end{enumerate}

\noindent At this point in time, we have already put some significant effort into (1):
Most importantly, we have implemented a new equivalence oracle, \textsc{AFLEQOracle}, in LearnLib, which iteratively loads a traces that AFL marks as interesting, and parses them as a test query for the learner.
Unfortunately, we were unable to apply this new equivalence oracle to the RERS challenge due to time restrictions.
The code for this project is available at \url{https://github.com/praseodym/learning-fuzzing}.

In this section we give an overview of our current effort on using mutation-based fuzzing as a source of counterexamples during learning.

\medskip

\begin{figure}[b]
	\centering
    \begin{tikzpicture}[>=stealth']
    \draw (0,0) rectangle (3,2);
    \node [below] at (1.5,2) {JVM};
    \draw (0.25,0.25) rectangle (2.75,1.5) node[midway] {LearnLib};
    \draw (4.5,0.25) rectangle (7,1.5) node [midway] {libafl};
    \draw (8.5,0) rectangle (11.5,2);
    \node [below] at (10,2) {AFL fork server};
    
	\draw (8.8,0.45) rectangle (11.05,1.45) [fill=white];
	\draw (8.9,0.35) rectangle (11.15,1.35) [fill=white];
    \draw (9,0.25) rectangle (11.25,1.25) [fill=white] node [midway] {Target process};
    \draw [<->] (3,0.75) -- (4.5,0.75) node[midway,fill=white] {JNI};
    \draw [<->] (7,0.5) -- (9,0.5) node[pos=0.4,fill=white] {queries};
    \draw [->] (8.65,0.5) -- (8.65,0.75) -- (8.9,0.75);
    \draw [->] (8.65,0.75) -- (8.65,1) -- (8.8,1);
    \draw [->] (7,1.25) -- (8.5,1.25) node[midway,fill=white] {setup};
    \end{tikzpicture}
	\caption{Architecture for combining LearnLib and AFL.}
	\label{fig:learner-architecture}
\end{figure}
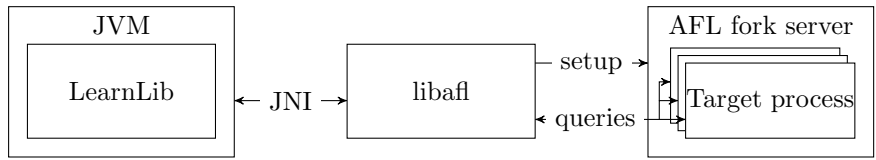

An overview of the architecture for combining AFL and LearnLib is shown in \autoref{fig:learner-architecture}.
To establish this, we had to tackle the following main issues:
\begin{itemize}
\item As AFL is provided as a standalone tool, we have created a library, libafl, that the learner can communicate with.
\item As LearnLib is written in Java, and AFL (and libafl) are written in C, we needed to bridge all communication between the two.
For this purpose, we have used the Java Native Interface (JNI) programming interface, which is part of the Java language.
JNI allows for code running in the Java Virtual Machine (i.e.\ LearnLib) to interface with platform-specific native binaries or external libraries (i.e.\ libafl).
\item We have added the possibility to embed the target program in AFL's fork server.
For each membership or test query, the fork server creates a new instance of the target process.
This speeds up the execution of learning, independent of the technique used to find counterexamples.
\end{itemize}

\noindent There were some other issues that we had to address:
\begin{itemize}
\item AFL is designed such that it does not care about the target program's output.
Instead only coverage data is used as a measure for test case relevancy. 
The learner, however, relies on output behaviour. 
Therefore, we have extended AFL to always save data from the target's \lstinline{stdout} into a shared memory buffer (shared between libafl and the fork server process). 
The content of this shared memory buffer is returned to LearnLib after a successful query.
\item AFL runs the target program in a non-interactive manner, i.e. it provides the program with input once and then expects it to terminate and reset state. 
This is in contrast to the default behaviour of LearnLib, which expects a \textit{single-step system under learning} that repeatedly accepts an input value and returns the associated output, and has an explicit option to reset.
We initially simulated this behaviour in AFL by running the target program once for each prefix of an input sequence.
For the RERS challenge, however, we could run each input sequence once, as it was easy to correlate individual inputs to their corresponding outputs.
\end{itemize}

\noindent We have performed some inital experiments with the setup described above.
In these experiments we compared different learning setups on their ability for finding error states in the reachability problems of the RERS 2015.
For these problems, the number of reachable error states are now known.

An selection of the results is shown in \autoref{tab:learning}.
In addition to the number of (reachability) states learned, this table compares learning performance in terms of learning time and the number of queries needed (lower is better).
In all cases, using fuzzing equivalence delivers models with more states and more reachability states found in a shorter learning time. 
One remark here is that the learning time we report only includes the time the learning process ran, not the time that the fuzzer ran. 
We ran the AFL fuzzer on each problem for one day, and the test cases that were generated during that time were used for equivalence testing using the learning process. 

\begin{table}
\centering
	\caption{Results for the RERS 2015 challenge problems on a Intel Xeon CPU E5-2430 v2 @ 2.50GHz (virtualised server), with Oracle Java 8 JVM configured with 4GB heap.}
	\label{tab:learning}
	\begin{tabular}{llllll}
	\hline
		{problem} 		& {method}				& {states}	& {errors} & {time} 	& {queries} \\               		
	\hline
        1		& TTT, W-method 1		& 25		& 19/29			& 4s & 7 342 \\
		1		& L*, W-method 8		& 25	    & 19/29			& 13h& $2.46 \times 10^8$ \\
   		1		& TTT, fuzzing			& 334		& 29/29			& 21s	& 16 731  	\\
		1		& \textbf{L*, fuzzing}	& 1 027		& 29/29 		& 44m	& $2.86 \times 10^6$ \\
	\hline
        2		& TTT, W-method 1		& 188		& 15/30			& 1h	& $8.15 \times 10^6$\\
		2		& L*, W-method 3		& 195	    & 15/30			& 17h	& $2.39 \times 10^7$ \\
   		2		& TTT, fuzzing			& 2 985		& 24/30			& 13m	& 412 340\\
		2		& \textbf{L*, fuzzing}	& 3 281		& 24/30			& 13h	& $4.21 \times 10^7$ \\
	\hline
		3		& L*, W-method 1		& 798	    & 16/32			& 110h	& $1.42 \times 10^9$ \\
   		3		& TTT, fuzzing			& 1 054		& 19/32			& 13m	& 698 409 \\
		3		& \textbf{L*, fuzzing}	& 1 094		& 19/32			& 13h	& $2.34 \times 10^7$ \\
	\hline
		4		& TTT, W-method 7		& 21	    & 1/23			& 4h	& $5.17 \times 10^7$ \\
   		4		& \textbf{TTT, fuzzing}	& 7 402		& 21/23			& 16m	& 458 763  \\
	\hline
		5		& L*, W-method 1		& 183       & 15/30			& 13h	& $2.20 \times 10^6$ \\
   		5		& \textbf{TTT, fuzzing}	& 3 376		& 24/30			& 8m	& 416 943 \\    
	\hline
		6		& L*, W-method 1		& 671       & 16/32			& 93h	& $8.89 \times 10^8$\\
   		6		& \textbf{TTT, fuzzing}	& 3 909		& 23/32			& 45m	& $1.80 \times 10^6$\\
	\hline
	\end{tabular}
\end{table}

\section{Conclusion}
An ongoing challenge for learning algorithms formulated in the Minimally Adequate Teacher framework is to efficiently obtain counterexamples.
In this paper we have compared and combined conformance testing and mutation-based fuzzing methods for obtaining counterexamples when learning finite state machine models for the reactive software systems of the RERS challenge.
We have found that for the LTL problems of the challenge the fuzzer did not find any additional counterexamples for the learner, compared to those found by the tester.
For the reachability problems of the challenge, however, the fuzzer discovered more reachable error states than the learner and tester, albeit in some cases the learner and tester found some that were not discovered by the fuzzer.
This leads us to believe that in some applications, fuzzing is a viable technique for finding additional counterexamples for a learning setup.


\begin{thebibliography}{12}
\providecommand{\natexlab}[1]{#1}
\providecommand{\url}[1]{\texttt{#1}}
\providecommand{\urlprefix}{}

\bibitem[{Angluin(1987)}]{angluin1987learning}
Angluin, D.: {Learning regular sets from queries and counterexamples}.
\newblock Information and Computation 75(2), 87--106 (1987)

\bibitem[{Dorofeeva et~al.(2010)Dorofeeva, El-Fakih, Maag, Cavalli, and
  Yevtushenko}]{Dorofeeva2010}
Dorofeeva, R., El-Fakih, K., Maag, S., Cavalli, A.R., Yevtushenko, N.:
  {FSM-based conformance testing methods: A survey annotated with experimental
  evaluation}.
\newblock Information and Software Technology 52(12), 1286--1297 (dec 2010),
  \urlprefix\url{http://linkinghub.elsevier.com/retrieve/pii/S0950584910001278}

\bibitem[{Fujiwara et~al.(1991)Fujiwara, Bochmann, Khendek, Amalou, and
  Ghedamsi}]{Fujiwara1991}
Fujiwara, S., Bochmann, G.V., Khendek, F., Amalou, M., Ghedamsi, A.: Test
  selection based on finite state models.
\newblock Software Engineering, IEEE Transactions on 17(6), 591--603 (1991)

\bibitem[{Isberner et~al.(2014)Isberner, Howar, and Steffen}]{Isberner2014a}
Isberner, M., Howar, F., Steffen, B.: {The TTT Algorithm: A Redundancy-Free
  Approach to Active Automata Learning}.
\newblock In: Proc.\ of RV, LNCS, vol. 8734, pp. 307--322 (2014)

\bibitem[{Merten et~al.(2011)Merten, Steffen, Howar, and
  Margaria}]{merten2011learnlib}
Merten, M., Steffen, B., Howar, F., Margaria, T.: {Next Generation LearnLib}.
\newblock In: Proc.\ of TACAS. LNCS, vol. 6605 (2011)

\bibitem[{Peled et~al.(1999)Peled, Vardi, and Yannakakis}]{Peled1999}
Peled, D., Vardi, M.Y., Yannakakis, M.: Formal Methods for Protocol Engineering
  and Distributed Systems, chap. Black Box Checking.
\newblock Springer US (1999)

\bibitem[{Rivest and Schapire(1994)}]{Rivest1994}
Rivest, R., Schapire, R.: {Diversity-based inference of finite automata}.
\newblock Journal of the ACM 41(3), 555--589 (1994)

\bibitem[{Smetsers et~al.(2016)Smetsers, Moerman, and Jansen}]{Smetsers2016}
Smetsers, R., Moerman, J., Jansen, D.N.: Minimal Separating Sequences for All
  Pairs of States, pp. 181--193.
\newblock Springer International Publishing, Cham (2016),
  \urlprefix\url{http://dx.doi.org/10.1007/978-3-319-30000-9_14}

\bibitem[{Vasilevskii(1973)}]{Vasilevskii1973}
Vasilevskii, M.P.: {Failure diagnosis of automata}.
\newblock Cybernetics 9(4), 653--665 (1973),
  \urlprefix\url{http://link.springer.com/10.1007/BF01068590}

\bibitem[{Zalewski(2014{\natexlab{a}})}]{afl-blog-forkserver}
Zalewski, M.: {Fuzzing random programs without execve()} (2014{\natexlab{a}}),
  \urlprefix\url{https://lcamtuf.blogspot.com/2014/10/fuzzing-binaries-without-execve.html},
  date accessed: 2015-09-15

\bibitem[{Zalewski(2015)}]{afl-website}
Zalewski, M.: {American Fuzzy Lop (AFL) fuzzer} (2015),
  \urlprefix\url{http://lcamtuf.coredump.cx/afl/}, date accessed: 2015-09-15

\bibitem[{Zalewski(2014{\natexlab{b}})}]{afl-blog-mutation}
Zalewski, M.: {Binary fuzzing strategies: what works, what doesn't}
  (2014{\natexlab{b}}),
  \urlprefix\url{https://lcamtuf.blogspot.nl/2014/08/binary-fuzzing-strategies-what-works.html}

\end{thebibliography}
\end{document}